\documentclass[reprint, twocolumn, amssymb, amsmath, bibnotes, aps, prl, groupedaddress,longbibliography]{revtex4-2} 

\usepackage{graphicx,comment}
\usepackage{float}
\usepackage[dvipsnames]{xcolor}
\usepackage[colorlinks=true]{hyperref}
\definecolor{darkblue}{RGB}{33,33,137}
\definecolor{darkred}{RGB}{193,23,23}
\hypersetup{
    colorlinks=true,       
    linkcolor=blue,     
    citecolor=blue,    
    urlcolor=blue      
}

\begin{document}

\title{Kibble--Zurek Meets Tricriticality: Breakdown of Adiabatic--Impulse and New Scaling Forms}

\author{Chengshu Li}
\email{chengshu@mail.tsinghua.edu.cn}
\affiliation{
Institute for Advanced Study, Tsinghua University, Beijing 100084, China
}

\date{\today}

\begin{abstract}
\noindent
\textbf{Keywords}: Kibble--Zurek effect, tricriticality, adiabatic--impulse scenario
\end{abstract}

\maketitle

The Kibble--Zurek (KZ) effect offers an overarching description of dynamical scaling behavior near a critical point~\cite{kibble1976topology,zurek1985cosmological}. Originally proposed in a classical setup, the KZ effect has been generalized to quantum phase transitions~\cite{zurek2005dynamics,dziarmaga2005dynamics,polkovnikov2005universal} and is actively explored on quantum simulation platforms~\cite{keesling2019quantum,ebadi2021quantum,manovitz2025quantum,zhang2025near}. Exploring how the KZ effect fares across different criticalities has proven to be a rewarding pursuit, significantly enriching our understanding of nonequilibrium quantum dynamics~\cite{zurek2005dynamics,dziarmaga2005dynamics,polkovnikov2005universal,Zhong2005,Bermudez2009Topology,bermudez2010dynamical,polkovnikov2011colloquium,Sandvik2011,kolodrubetz2012nonequilibrium,Sondhi2012,kolodrubetz2012nonequilibrium,del2014universality,Yin2017,ChenInsulatorKZ2020,rossini2021coherent,Nie2024topo,shu2025equilibration,Deng2025}.

Recently, in a work published in Chinese Physics Letters, Wang et al.~\cite{yin2025driven} studied the KZ effect near a tricritical Ising (TCI) phase transition point, and brought many new insights into this vibrant research area~\footnote{For a concurrent work discussing similar physics, see \cite{wang2025}.}. The main discoveries are twofold. First, despite that the established prerequisite of ``adiabatic--impulse scenario'' (AIS) for KZ breaks down near TCI, the KZ scaling remains intact. This establishes KZ as a broader framework than AIS in a concrete, real-time setting~\cite{Zeng2025}. Second, combining physical arguments and numerical evidence, the authors propose a novel dynamical scaling law for the entanglement entropy, with a surprising combination of two central charges within a single equation.

More concretely, the authors focus on the one-dimensional Grover--Sheng--Vishwanath model~\cite{grover2014emergent,Li2020TCI}, with the Hamiltonian
\begin{equation}
\begin{split}
H=&-i\sum_j[1-g\mu_{j+1/2}^z+(-1)^j(h_\sigma-1)]\chi_j\chi_{j+1}\\
&+\sum_j\mu_{j-1/2}^z\mu_{j+1/2}^z-h_\mu\sum_j\mu_{j+1/2}^x.    
\end{split}
\end{equation}
The model is composed of a Majorana chain of $\chi$ and a transverse field Ising chain of $\mu$. The order parameter of the latter $(-1)^j\mu^z$ results in a dynamical mass for the former, and one also includes a static mass $h_\sigma-1$. Hence, for $h_\sigma=1$ and large $h_\mu$, the Majorana chain is massless and the low energy physics is described by a $c=1/2$ Ising conformal field theory (CFT), with $c$ the central charge. If one keeps $h_\sigma=1$ but considers small $h_\mu$, the $\mu$-spin chain spontaneously breaks its antiferromagnetic $\mathbb{Z}_2$ symmetry and opens up a gap for the $\chi$-chain. As a generic scenario, a TCI phase transition lies in between the $c=1/2$ phase and the gapped phase, with a distinct central charge $c=7/10$.

The authors consider a ramping protocol from large $h_\mu$ across the TCI point, with the initial state the ground state of the starting point. Because the initial ramping stage stays exclusively in the parameter regime where the Hamiltonian is gapless, the usual argument of AIS based on adiabaticity does not make sense. Nevertheless, thanks to the fact that $z'<z+1/\nu_\mu$, where $z=1$ ($z'=1$) is the dynamic exponent of the TCI (Ising) CFT and $\nu_\mu=5/4$ is the correlation length exponent of TCI related to $h_\mu$, one can still imagine that the ramping is ``slow'' compared to the intrinsic dynamics far away from TCI. Applying the KZ equation, one expects the scaling forms for bosonic and fermionic correlation functions,
\begin{align}
&M_2= L^{-\eta_b}\mathcal{F}_1(RL^{r_\mu}),\\
&C_f= L^{-1-\eta_f}\mathcal{F}_3(RL^{r_\mu}),\label{eq:fermion1}
\end{align}
where the correlation functions are defined by $M_2=(1/L)\sum_i\langle\mu_0^z\mu_i^z\rangle$, $C_f=(1/L)\sum_i\langle\chi_i\chi_{i+L/2}\rangle$, $L$ is the system size, $\eta_b=\eta_f=2/5$ are the anomalous bosonic and fermionic dimensions (they are equal because of emergent supersymmetry at TCI~\cite{Friedan1985}), and $r_\mu=z+1/\nu_\mu=9/5$ is the all-important KZ exponent related to the ramping rate $R$. Furthermore, at large $R$, one expects the correlation functions to recover the parent Ising form, hence
\begin{align}
&M_2= L^{-1}R^{-(1-\eta_b)/r_\mu},\\
&C_f= L^{-1}R^{\eta_f/r_\mu},\label{eq:fermion2}
\end{align}
to which we will return soon. All these scaling forms are corroborated by the numerical simulations reported in \cite{yin2025driven}.

The authors then consider how the half-system entanglement entropy $S$ scales with $R$. By matching the TCI and Ising scaling forms, and benchmarking against numerics, the authors propose the following large-$R$ result: 
\begin{equation}
S=\frac{c_\mathrm I}{3}\log L+\frac{c_\mathrm I- c_\mathrm T}{3r_\mu}\log R,\label{eq:S}
\end{equation}
where $c_\mathrm I$ ($c_\mathrm T$) is the Ising (TCI) central charge.
First, it is not hard to see that it satisfies the TCI scaling form 
\begin{equation}
S=\frac{c_\mathrm T}{3}\log L+\mathcal{F}_4(RL^{r_\mu}).
\end{equation}
Second, when fixing (large) $R$, one also recovers the Ising result. This equation hence gives a novel description of the universal scaling behavior of the entanglement entropy.

The outstanding feature of Eq.~\eqref{eq:S} is that two central charges from two different CFTs appear together in a single equation. It is natural to ask whether such a feature can be recovered in other critical exponents. Indeed, \`a la Calabrese--Cardy~\cite{Calabrese-Cardy_2004}, the entanglement entropy is given by the (logarithmic) correlation function of twist operators, and the central charge appears in the scaling dimension of these operators. Returning to the usual fermionic correlation function Eq.~\eqref{eq:fermion1}, although the authors did not make the comparison explicit, it is perhaps instructive to rewrite it as
\begin{equation}
\log C_f=-(1+\eta_f^\mathrm T)\log L+\log\mathcal{F}_3(RL^{r_\mu}),
\end{equation}
where we have added the superscript T to emphasize that the exponent comes from TCI.
In particular, for large $R$, one has
\begin{equation}
\log C_f=-(1+\eta_f^\mathrm I)\log L+\frac{\eta_f^\mathrm T-\eta_f^\mathrm I}{r_\mu}\log R.\label{eq:fermion3}
\end{equation}
This form makes the parallel between Eqs.~\eqref{eq:S} and \eqref{eq:fermion2} clearest. Note that $\eta_f^\mathrm I=0$ for free fermions.

These intriguing results naturally lead to many opportunities for further research of nonequilibrium quantum dynamics. The mixing of different central charges is reminiscent of Zamolodchikov's celebrated $c$-theorem~\cite{zamolodchikov1986irreversibility}. Generalizing the $c$-theorem to a dynamical setting~\cite{wang2025} is a promising direction that might include the current results as an illuminating example. While in Eq.~\eqref{eq:fermion3} $\eta_f^\mathrm I=0$ for the Ising CFT, this is generally not the case and calls for further investigation in, say, the tricritical Potts CFT. Finally, based on recent proposals to realize TCI in a Rydberg quantum simulator~\cite{li2024supersymmetry,wang2025}, it is promising that the predictions in this work be experimentally tested in the near future.

\bibliography{Rydberg}

\end{document}